\documentstyle[12pt,epsf]{article}
\newcommand{\beq}{\begin{equation}}
\newcommand{\eeq}{\end{equation}}
\newcommand{\beqa}{\begin{eqnarray}}
\newcommand{\eeqa}{\end{eqnarray}}
\newcommand{\boldtau}{\mbox{\boldmath $\tau$}}

\newcommand{\boldpi}{\mbox{\boldmath $\pi$}}

\begin{document}

\begin{titlepage}

\hfill{KRL MAP-269}

\vspace{2.5cm}

\begin{center}
{\Large\bf The Nucleon Anapole Form Factor \\
in Chiral Perturbation Theory to Sub-leading Order}

\vspace{2.0cm}

{\bf C.M. Maekawa}\footnote{{\tt maekawa@krl.caltech.edu}},
 {\bf J. S. Veiga}\footnote{{\tt jaime@krl.caltech.edu}}
and 
{\bf U. van Kolck}\footnote{{\tt vankolck@krl.caltech.edu}}

\vspace{0.8cm}
{\it
 Kellogg Radiation Laboratory, 106-38 \\
 California Institute of Technology \\
 Pasadena, CA 91125}
\end{center}

\vspace{1.5cm}

\begin{abstract}
The anapole form factor of the nucleon is
calculated in chiral perturbation theory to sub-leading order.
This is the lowest order in which the  
isovector anapole form factor does not vanish.
The anapole moment depends on counterterms that reflect
short-range dynamics, but the momentum dependence of
the form factor is determined by pion loops in terms of parameters
that could in principle be fixed from other processes.
If these parameters are assumed to have natural size,
the sub-leading corrections do not exceed $\sim$ 30\%
at momentum $Q\sim 300$ MeV.

\end{abstract}

\vspace{2cm}
\vfill
\end{titlepage}

\setcounter{page}{1}

Parity-violating electron scattering 
has long played a role in understanding electroweak
interactions, and has more recently been explored as a tool
for the study of
nucleon structure.
The SAMPLE collaboration has carried out electron scattering measurements at
a momentum transferred of $Q^2=0.1$ MeV$^2$ on both the
proton \cite{Sample} and the deuteron \cite{SampleII},
for a simultaneous extraction of
the strange magnetic ($G_{M}^{s}$) and 
the axial form factor of the nucleon ($G_{A}^{e}$). 

One quantity that contributes in electron scattering as $G_{A}^{e}$
is the anapole form factor, 
which is an extension for $Q^{2}> 0$ of the anapole moment. 
The anapole is a parity-violating electromagnetic moment of 
a charge particle with spin \cite{Zei58}.
Recently the effect of the nuclear anapole moment 
in atomic parity violation was measured 
precisely in $^{133}$Cs transitions \cite{Wood97}, 
and a discrepancy with theory found.
Parity violation in this case
is enhanced 
by nuclear medium effects.
No such enhancement is present
in parity-violating electron scattering off the proton and deuteron; 
however, the anapole form factor could still be visible. 
Using previous estimates of the anapole moment \cite{Mus+91,Sav+93},
the proton data 
implies a positive value for $G_{M}^{s}$ \cite{Sample}, 
in disagreement with most theoretical predictions 
(for a summary, see Ref. \cite{Bob}).

Experiments of current interest \cite{Sample,SampleII,Happex,G0,tito}
are performed at finite $Q^2=-q^2$.
For $Q<M_{QCD}$, where $M_{QCD}\sim 1$ GeV is the characteristic QCD mass 
scale, we are deep in the non-perturbative regime of QCD,
where currently the only possible systematic calculations
are in terms of hadrons.
At $Q\sim O(m_{\pi })$  the photon can resolve the pion cloud around 
the non-relativistic nucleon,
and calculations are possible in  Chiral Pertubation 
Theory (ChPT), which involves pions, nucleons,
and delta isobars, and which has been successfully applied to hadronic
and nuclear systems 
\cite{mei,bira}.
The first anapole calculations were limited 
to $Q^{2}=0$ in leading \cite{Mus+91,Sav+93} and sub-leading
orders \cite{Sav+93,mus+00}.
Recently, the full form factor of the nucleon was calculated
in leading order \cite{Sav+98,bira1}.
In this order the form factor comes entirely from
the pion cloud and is purely
isoscalar, while experiments are most sensitive
to the isovector component \cite{Bob}.
Here we report results of sub-leading contributions to the
nuclear anapole form factor, where the isovector part
first appears.

In the framework of ChPT, QCD symmetries are used as a guide to 
build the most general effective 
Lagrangian. The number of terms in the Lagrangian is not constrained by 
symmetries, which demands
a power counting argument to order interactions
according to the expected size of their contributions. 
In order to fullfill chiral symmetry requirements, 
pions couple derivatively in the chiral limit; this
derivative coupling brings to the 
amplitude powers of pion momentum or
powers of the delta-nucleon mass difference 
(comparable to the pion mass). 
Chiral symmetry breaking terms involve quark masses, so they 
bring into the amplitude powers of the pion mass. 
Thus one has a chiral 
index ($\Delta $) available to order the Lagrangian terms, 
${\cal L}=\sum_{\Delta }{\cal L}^{\left( \Delta \right) }$.
For strong interactions, the index counts
powers of $Q/M_{QCD}$, and it is given by 
$\Delta =d+n/2-2$, 
where $n$ is the number of fermions fields and $d$ counts the 
numbers of derivatives, powers of the pion mass, and of the 
delta-nucleon mass difference.
In the presence of electromagnetic interactions,
it is convenient to include in $d$ powers of the charge $e$ as well.
Weak interactions, on the other hand, bring powers of a very small
factor $G_F f_\pi^2$, where 
$G_F$ is the Fermi constant and $f_\pi$ the pion decay constant.
Since we count these factors explicitly, negative indices appear.
 
Based on this power counting argument the interactions relevant to our
problem are the following.
The parity-conserving terms are well known \cite{mei}:
\begin{eqnarray} 
{\cal L}_{str/em}^{( 0) }&=&
         \frac{1}{2}\left( D_{\mu }{\boldpi }\right) ^{2}
        -\frac{1}{2}m_{\pi }^{2}{\boldpi }^{2}+\bar{N}iv\cdot DN
        -\frac{g_{A}}{f_{\pi }}\bar{N}
         \left( {\boldtau }\cdot S\cdot D{\boldpi }\right)N
        +\ldots \label{Lstr0}\\
{\cal L}_{str/em}^{( 1) }&=& 
        \frac{1}{4m_{N}}\bar{N}\left[ (v\cdot D)^{2}-D^{2}\right]N
        + i\frac{g_A}{2m_N f_\pi}
         \bar{N}\left\{S\cdot D,\boldtau\cdot v \cdot D \boldpi \right\} N
         \nonumber \\ 
&&-\frac{i}{4m_{N}}
   \bar{N}\left[ S^{\mu},S^{\nu}\right]
   \left[1+\kappa^{s}+(1+\kappa^{v})\tau_{3}\right]N F_{\mu\nu}+ \ldots
\label{Lstr1}
\end{eqnarray}
Here $\boldpi$ denotes the pion field with
$f_{\pi }=93$ MeV the pion decay constant;
$N$ represents the heavy nucleon field of 
four-velocity $v^{\mu }$ and spin $S^{\mu }$
(in the nucleon rest frame $v^{\mu }=(1,\vec{0})$ and
$S^{\mu }=( 0,\vec{\sigma}/2)$);
$A_{\mu }$ is the photon field and $F_{\mu\nu}$ is the photon strength field;
$D_{\mu }=( \partial _{\mu}-ieQA_{\mu })$ is the covariant
derivative, with 
$Q_{ab}^{( \pi) }=-i\varepsilon_{3ab}$ for a pion
and $Q^{( N) }=( 1+\tau _{3}) /2$ for a nucleon;
and ``...'' stands for other interactions with more pions, nucleons 
and deltas.
The pion-nucleon coupling $g_{A}$ and the magnetic photon-nucleon couplings
 $\kappa^{(s)}$ and $\kappa^{(v)}$ are not determined
from symmetry but expected to be $O(1)$; indeed, one finds
$g_{A}=1.267$,
$\kappa^{(s)}=-0.12$, and $\kappa^{(v)}=5.62$ \cite{mei}.

The relevant parity-violating terms were discussed in Ref. \cite{Sav+93}:
\begin{eqnarray}
{\cal L}_{weak}^{( -1) } &=&
                 -\frac{h^{(1)}_{\pi NN}}{\sqrt{2}}
                  \bar{N}( {\boldtau }\times {\boldpi } )_{3} N
                 + \ldots  \label{Lw-1} \\
{\cal L}_{weak}^{( 0) } &=&
                 -\frac{2}{f_{\pi }^{2}}
                  \bar{N}S^{\mu}
                  \left\{ \left( h_{A}^{(1)}+h_{A}^{(2)}\tau _{3}\right) 
                  \left[ \left( \boldpi\times \partial _{\mu }\boldpi\right)_{3} 
               +eA_{\mu }\left( \boldpi ^{2}-\pi_{3}^{2}\right) \right] \right.
                \nonumber \\
&& \qquad \qquad \qquad +\left. h_{A}^{(2)}(\boldpi \times \boldtau )_{3}
             \partial _{\mu }\pi_{3}\right\}N  \nonumber \\
&& +\frac{1}{f_{\pi }}\bar{N}
\left[ \left( h_{V}^{(0)}+\frac{4}{3}h_{V}^{(2)}\right)\frac{1}{2}\boldtau\cdot v\cdot D\boldpi
                     -2h_{V}^{(2)}\tau_3v\cdot D\pi _{3}\right]N  
+ \ldots \label{Lw0vec} \\
{\cal L}_{weak}^{( 2) } &=&
                \frac{2}{m_{N}^{2}}\bar{N}
                \left(\tilde{a}_{0}+\tilde{a}_{1}\tau _{3}\right) S_{\mu }N
                \partial _{\nu }F^{\mu\nu }+\ldots  \label{Lw2}
\end{eqnarray}
Here $h_{\pi NN}^{(1)}$, $h_{A}^{(1,2)}$ and $h_{V}^{(0,2)}$ are, respectively,
Yukawa, axial-vector and vector parity-violating 
pion-nucleon couplings,
with superscripts refering to isospin $\Delta I=0,1$ and 2.
On the basis of naive dimensional analysis,
$h_{\pi NN}^{(1)}f_{\pi }\sim O(G_Ff_{\pi }^2 M_{QCD})$, and
$h_{A}^{(1,2)}\sim h_V^{(0,2)} \sim O(G_Ff_{\pi }^2 )$.
Also, $\tilde{a}_{0,1}$ are short-range contributions
to the anapole moment, expected to be of $O(eG_F f^2_\pi /m^2_N)=
O(eG_F /(4\pi)^2)$.

The current-current electron-nucleon interaction has the form 
\begin{equation}
iT=-ie\bar{e}\left( k^{\prime }\right) \gamma ^{\mu }e\left( k\right)
D_{\mu \nu }\left( q\right) \bar{N}\left( p^{\prime }\right) 
J_{an}^{\nu}\left( q\right) N\left( p\right),
\end{equation}
where $e( k) $ ($N( p) $) is an electron 
(nucleon) spinors of momentum $k$ ($p$), 
$-e$ is the electron charge, 
$iD_{\mu \nu}( q) =-i \eta _{\mu \nu }/q^{2} $ is the photon
propagator with $q^{2}=( p-p^{\prime }) ^{2}\equiv -Q^{2}<0$, and
the nucleon anapole current $iJ_{an}^{\mu }$ reads
\begin{equation}
J_{an}^{\mu }( q) =\frac{2}{m_{N}^{2}}
                   \left[ a_{0}F_{A}^{(0) }(-q^{2}) 
                         +a_{1}F_{A}^{(1) }(-q^{2}) \tau _{3}\right] 
                   \left( S^{\mu }q^{2}-S\cdot qq^{\mu }\right),
\end{equation}
where $a_{0}$ and $ a_{1}$ are the
isoscalar and isovector anapole
moments, and $F_{A}^{( 0) }( -q^{2}) $ and 
$F_{A}^{( 1) }( -q^{2}) $ their corresponding
form factors.
 
\begin{figure}[tb]
\begin{center}
\epsfxsize=8cm
\centerline
{\epsffile{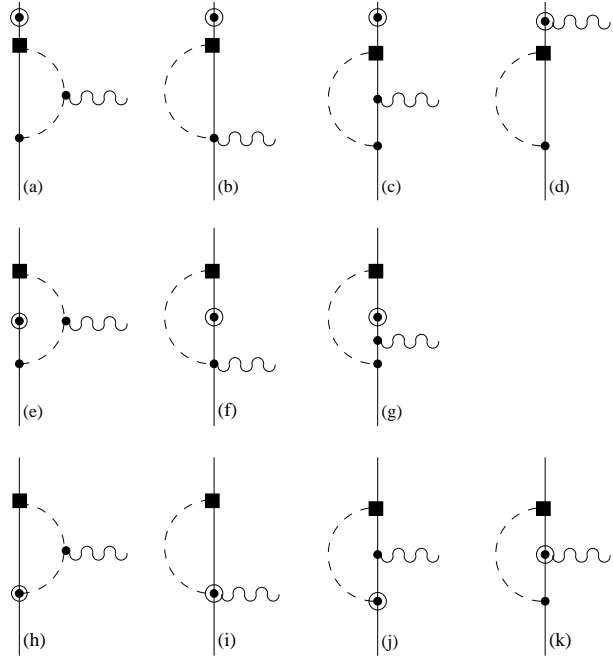}}
\end{center}
\caption{Diagrams contributing to the nucleon anapole form factor 
in sub-leading order 
coming from one insertion of an ${\cal L}_{str/em}^{( 1) }$ operator.
Solid, dashed and wavy lines represent nucleon, 
pions and (virtual) photons, respectively; 
squares represent the parity-violating vertex
from ${\cal L}^{(-1)}_{weak}$; 
single filled circles stand for interactions from 
${\cal L}^{(0)}_{str/em}$
and double circles represent interactions from 
${\cal L}^{(1)}_{str/em}$. 
For simplicity only one possible orderings are shown 
here.}
\label{figure1}
\end{figure}

\begin{figure}[tb]
\begin{center}
\epsfxsize=6cm
\centerline
{\epsffile{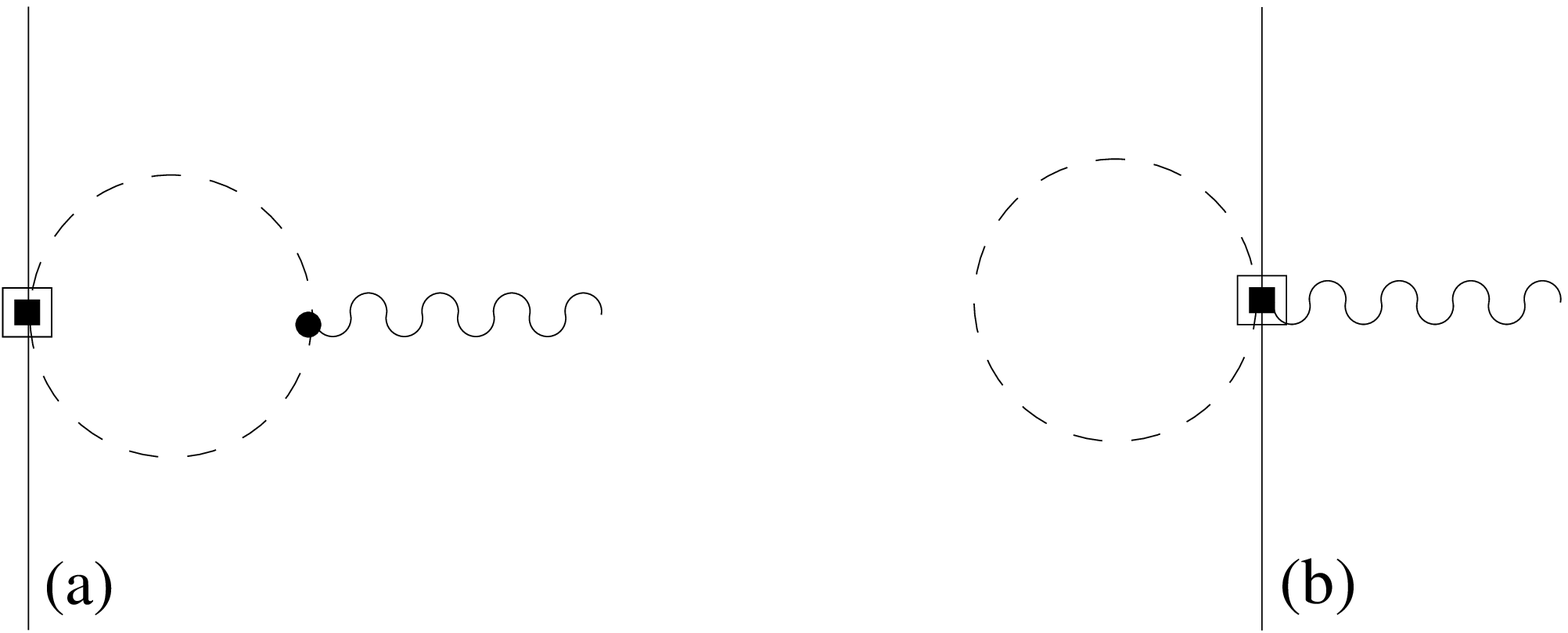}}
\end{center}
\caption{Diagrams contributing to the nucleon anapole form factor 
in sub-leading order 
coming from one insertion of the axial-vector
couplings in ${\cal L}_{weak}^{(0) }$,
represented by a double square. 
Other symbols are as in Fig. \ref{figure1}.}
\label{figure2}
\end{figure}

\begin{figure}[tb]
\begin{center}
\epsfxsize=6cm
\centerline
{\epsffile{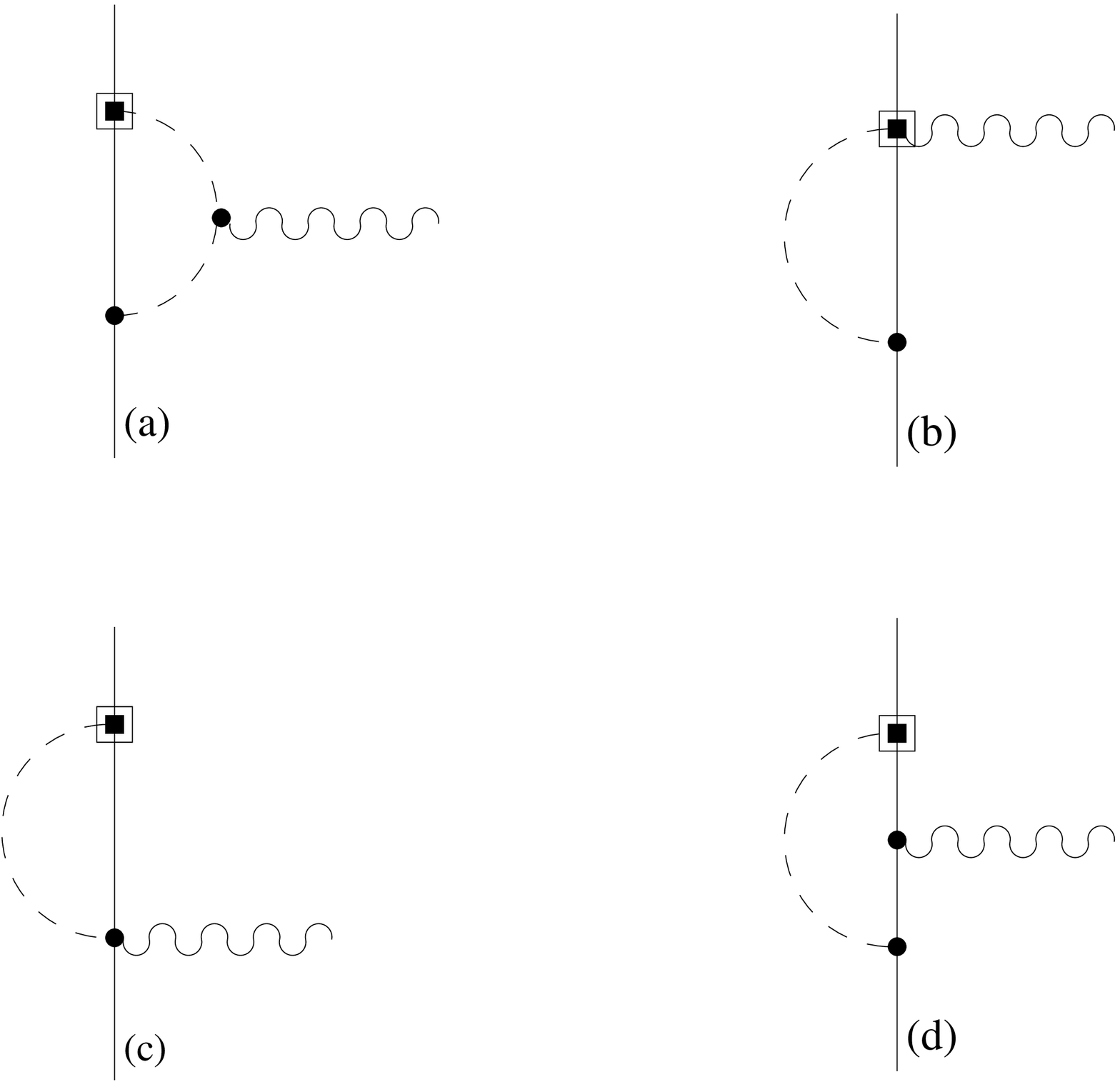}}
\end{center}
\caption{Diagrams contributing to the nucleon anapole form factor 
in sub-leading order 
coming from one insertion of the vector couplings in ${\cal L}_{weak}^{(0) }$,
represented by a double square.
Other symbols are as in Fig. \ref{figure1}.
For simplicity only one of two possible 
orderings are shown here.}
\label{figure3}
\end{figure}

The diagrams contributing to the nucleon anapole form factor 
in next-to-leading order 
(NLO) are shown in Figs. \ref{figure1},\ref{figure2},\ref{figure3}.
We classify them according to the combination of couplings 
that appear.

The NLO diagrams of  Fig. \ref{figure1} are built from
the leading interactions in
${\cal L}_{str/em}^{( 0) }$ and ${\cal L}_{weak}^{( -1) }$, plus 
one insertion of an operator from ${\cal L}_{str/em}^{( 1) }$.
This insertion can be (i) a kinetic correction 
---either
in the nucleon propagator or in the external energy--- to 
the leading order (LO) diagrams 
computed in Ref. \cite{bira1};
or (ii) a sub-leading (magnetic) photon-nucleon interaction.
The size of these diagrams is
$O(eG_FQ^2/(4\pi)^2)$.
Indeed, LO
contributions are $O(eG_FM_{QCD}Q/(4\pi)^2)$ \cite{bira1},
and NLO is of relative size $O(Q/M_{QCD})$.
(For example, the diagram $\ref{figure1}e$ has a kinetic
insertion of $Q^2/m_N$ and an extra propagator $1/Q$
compared to the corresponding LO diagram.)
Diagrams (c), (g) and (j) do not contribute to the 
anapole form factor because they are proportional 
to $v^{\mu}$ and the diagram (d) vanishes because
it is proportional to $S\cdot v =0$. 
Diagram (i) gives a pure isovector contribution, but it gets cancelled
by the isovector part of diagram (k). 
Therefore, the sum of all 
diagrams in Fig. \ref{figure1} is a purely isoscalar result.

The diagrams in Fig. \ref{figure2} 
have axial-vector vertices from ${\cal L}_{weak}^{( 0) }$.
They have both isovector and isoscalar parts. 
To 
evaluate the size of the contributions represented by these diagrams, 
one takes, for example, 
the diagram 2a: it has a parity-violating
two pion-nucleon axial vertex of the order $G_FQ$, 
a photon-pion 
vertex of $O(eQ)$, two pion propagators each one of $O(1/Q^2)$, 
and the loop integration 
of $O(Q^4/(4\pi^2))$. 
Diagrams of this type are then also of $O(eG_F Q^2/(4\pi)^2)$.

In Fig. \ref{figure3},
diagrams contain vector couplings coming from ${\cal L}_{weak}^{( 0) }$.  
Since the parity-violating vector coupling is $O(Q/M_{QCD})$ 
smaller than the LO Yukawa coupling, these contributions are 
clearly also $O(eG_F Q^2/(4\pi)^2)$.
Diagrams (b) and (d) are proportional to $v^{\mu}$ and do not 
contribute to the anapole form factor. 
Diagrams (a) and (c) give a purely isovector contribution.

\begin{figure}[tb]
\begin{center}
\epsfxsize=2cm
\centerline
{\epsffile{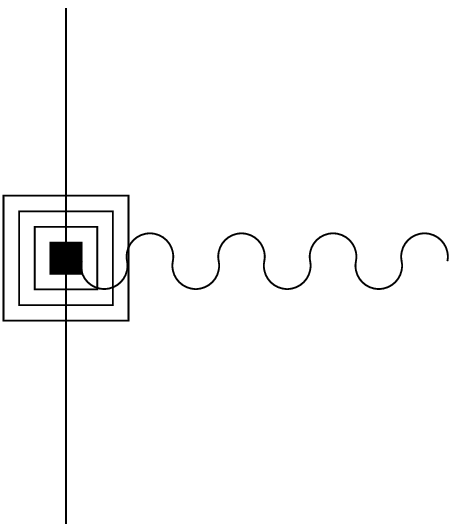}}
\end{center}
\caption{Diagram contributing to the nucleon anapole moment
in sub-leading order 
coming from ${\cal L}_{weak}^{(2) }$,
represented by a quadruple square.
Other symbols are as in Fig. \ref{figure1}.}
\label{figure4}
\end{figure}

Finally, there are short range contributions from ${\cal L}_{weak}^{( 2) }$
depicted in Fig. \ref{figure4}.
{}From the size of $\tilde{a}_{0,1}$, we see that these contributions are also
$O(eG_F Q^2/(4\pi)^2)$.

Note that to this order there are no contributions from the delta
isobar \cite{mus+00}.
Deltas would contribute at this order through intermediate states of
diagrams with one pion loop, e.g. diagram \ref{figure1}c with one
nucleon propagator replaced by a delta:
at least there would be one $\pi N \Delta$ 
vertex, either parity conserving or violating, and both 
kinds of vertices have the same $i\gamma_5$ structure,
which vanishes  in the framework of ChPT.
The first non-vanishing delta contribution shows up in an order higher than we
are considering here.

Let us first discuss the isoscalar component,
which did not vanish in leading order \cite{bira1},
\begin{equation}
a_{0}^{LO}=
\frac{eg_{A}h_{\pi NN}^{(1)}}{48\sqrt{2}\pi f_{\pi } }\frac{m_{N}^2}{m_{\pi }}.
\end{equation}
As the final contribution represented by the diagrams in 
Fig. \ref{figure1} is isoscalar, we add it 
to the isoscalar contribution 
of the diagrams in Fig.  \ref{figure2}, and find for
the anapole moment in next-to-leading order,
\begin{equation}
a_{0}^{NLO}=\tilde{a}_{0}( \mu) 
            +\frac{em_{N}^2}{(4\pi)^{2}f_{\pi }^{2}}
             \left( -\frac{g_{A}h_{\pi NN}^{(1)}f_{\pi }}{\sqrt{2}m_{N}}
            +\frac{h_{A}^{(1)}}{3}\right) 
            \ln \left( \frac{\mu ^{2}}{m_{\pi }^{2}}\right),
\label{NLOscalarmom}
\end{equation}
with
\begin{eqnarray}
\tilde{a}_{0}( \mu) &=&\tilde{a}_0
                       +\frac{em_{N}^2}{(4\pi)^{2}f_{\pi }^{2}}
      \left[ \left( -\frac{g_{A}h_{\pi NN}^{(1)}f_{\pi }}{\sqrt{2}m_{N}}
      +\frac{h_{A}^{(1)}}{3}\right) 
      \left( \frac{1}{\varepsilon }+1-\gamma+\ln 4\pi \right) \right. 
\nonumber\\
& &\qquad \qquad \qquad- \left. \frac{1}{3}
     \left(-\frac{g_{A}h_{\pi NN}^{(1)}f_{\pi }}{\sqrt{2}m_{N}}
           +\frac{2h_{A}^{(1)}}{3}\right) \right] ,
\label{scalarmommu}
\end{eqnarray}
where $\mu $ is the renormalization scale and $\gamma=0.5772157$ is the Euler constant.
As usual in ChPT, 
the only term that can be calculated explicitly 
is non-analytic in the pion mass; it
has the expected size, that is, it is $O(m_\pi/M_{QCD})$
smaller than $a_{0}^{LO}$.
This result for the anapole moment
agrees with that of a previous calculation \cite{mus+00}.
The term in $h_{A}^{(1)}$ agrees with Ref. \cite{Sav+93}.

The total isoscalar form factor reads
\begin{eqnarray}
F_{0}^{LO+NLO}( Q^{2}) &=& 1
        +\frac{a_{0}^{LO}}{(a_{0}^{LO}+a_{0}^{NLO})}
         \left[ F_{0}^{LO}( Q^{2})-1\right] \nonumber \\
       &&+\frac{1}{( a_{0}^{LO}+a_{0}^{NLO}) }
         \frac{em_{N}^2}{3(4\pi f_{\pi})^{2}}
         \left\{ \frac{g_{A}h_{\pi NN}^{(1)}f_{\pi }}{\sqrt{2}m_{N}}
         \left[F^{NLO1}( Q^{2}) -1\right] \right. \nonumber \\
        && \qquad \qquad \qquad \qquad \qquad \qquad%
       \left. -\frac{2h_{A}^{(1)}}{3}\left[F^{NLO2}( Q^{2}) -1\right] \right\}   
\end{eqnarray}
where $F_{0}^{LO}( Q^{2}) $ is the leading-order 
form factor given by \cite{bira1} 
\begin{equation}
F_{0}^{LO}( Q^{2}) =\frac{3}{2}
            \left\{ -\left( \frac{2m_{\pi } }{\sqrt{Q^{2}}}\right) ^{2}
      +\left[ \left( \frac{2m_{\pi } }{\sqrt{Q^{2}}}\right)^{2}+1\right] 
        \frac{2m_{\pi } }{\sqrt{Q^{2}}}
        \arctan \frac{\sqrt{Q^{2}}}{2m_{\pi }}\right\},
\end{equation}
$F^{NLO1}( Q^{2})$ comes from 
the diagrams in Fig. \ref{figure1},
\begin{equation}
F^{NLO1}( Q^{2}) =-3\left[\left(\frac{2m_{\pi }}{Q}\right)^2+2\right]
                        \left[ 1-
   \frac{1}{2}\sqrt{1+\left(\frac{2m_{\pi }}{Q}\right)^2}
           \ln \frac{\sqrt{1+\left(\frac{2m_\pi}{Q}\right)^2}+1}
                    {\sqrt{1+\left(\frac{2m_\pi}{Q}\right)^2}-1}\right] ,
\end{equation}
and 
$F^{NLO2}( Q^{2})$ comes from 
the diagrams in Fig. \ref{figure2},
\begin{equation}
F^{NLO2}( Q^{2})=-3\left[\left(\frac{2m_\pi}{Q}\right)^2+1\right] 
                      \left[ 1-
   \frac{1}{2}\sqrt{1+\left(\frac{2m_\pi}{Q}\right)^2}
           \ln \frac{\sqrt{1+\left(\frac{2m_\pi}{Q}\right)^2}+1}
                    {\sqrt{1+\left(\frac{2m_\pi}{Q}\right)^2}-1}\right].
\end{equation}

\begin{figure}[tb] 
\begin{center}
\epsfxsize=10cm
\centerline
{\epsffile{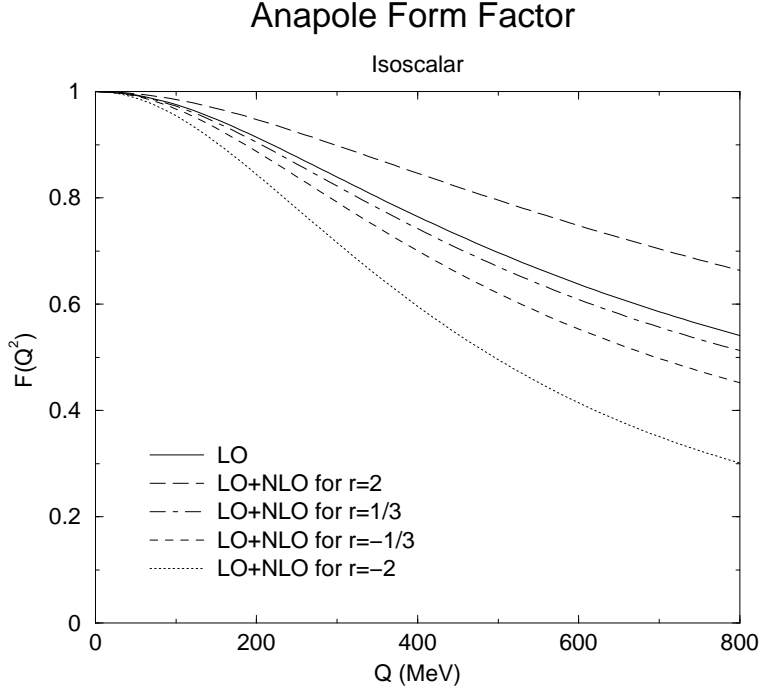}} 
\end{center}
\caption{The isoscalar anapole form factor $F_0$
as function of $Q$ in $\chi$PT: leading order (LO) and
next-to-leading order (NLO)
for a few reasonable values of parameters
expressed by the ratio 
$r=\sqrt{2}m_{N}h_{A}^{1}/3g_{A}f_{\pi } h_{\pi NN}^{1}$ .}
\label{isoscalar}
\end{figure}

As in lowest order, the momentum dependence is fixed
by the pion cloud, and therefore the scale for
momentum variation is determined by $2m_\pi$.
Because there are several contributions to the form factor,
for which we follow the conventional normalization to 1, 
the exact form depends also
on the coupling constants that contribute to the anapole moment.
Unfortunately these are currently not well determined
by other data; once they are, one can plot the form
to this order.
Here we can only study ``reasonable'' estimates of
the momentum dependence.
Assuming \cite{Sav+93} $\tilde{a}_{0}(\Lambda _{\chi SB}) =0$ 
where $\Lambda _{\chi SB} \sim 4\pi f_\pi$ is 
the chiral symmetry breaking scale,
we rewrite 
\begin{eqnarray}
F_{0}^{LO+NLO}( Q^{2}) &\simeq&F_{0}^{LO}( Q^{2}) 
                          +\frac{3m_{\pi }}{\pi m_N} \left( 1-r\right) 
               \ln \left( \frac{\Lambda _{\chi SB}^{2}}{m_{\pi }^{2}}\right) 
                \left[ F_{0}^{LO}( Q^{2}) -1\right] \nonumber \\
&&\qquad + \frac{m_{\pi }}{\pi m_N}
           \left\{ \left[ F^{NLO1}( Q^{2})-1\right] 
                  -2r\left[ F^{NLO2}( Q^{2}) -1\right] \right\}, 
\end{eqnarray}
where $r=\sqrt{2}m_{N}h_{A}^{(1)}/3g_{A}f_{\pi }h_{\pi NN}^{(1)}\sim 1/3$. 
In Fig. \ref{isoscalar} we show 
$F_{0}^{LO}( Q^{2})$ and $F_{0}^{LO+NLO}( Q^{2}) $ 
for several values of $r$.

{}From the form factor is easy to extract closed forms for the
mean square radius.
We find
\begin{equation}
\left\langle r_{0}^{2}\right\rangle ^{LO+NLO}=\frac{3}{10m_{\pi }^{2}}
   \frac{1}{( a_{0}^{LO}+{a}_{0}^{NLO}) }
   \left[ a_{0}^{LO}-\frac{2em_N^{2}}{3(4\pi f_{\pi})^{2}}
   \left( \frac{4g_{A}h_{\pi NN}^{(1)}f_{\pi }}{\sqrt{2}m_N}-h_{A}^{(1)}\right) 
   \right],
\end{equation}
Using the same estimates as for the form factor,
\begin{equation}
\left\langle r_{0}^{2}\right\rangle ^{LO+NLO}\simeq 
   \frac{3}{10m_{\pi}^{2}}\left[ 1+\frac{6m_{\pi }}{\pi m_N}\left( 1-r\right)
    \left( \ln \frac{\Lambda _{\chi SB}}{m_{\pi }}-1\right) 
   -\frac{2m_{\pi }}{\pi m_N}\right].
\end{equation}
For $r$ ranging from $-2$ to 2, 
$\left\langle r_{0}^{2}\right\rangle ^{LO+NLO}$ ranges from 
3 to $1\times 10^{-5}$ MeV$^{-2}$.

The isovector anapole moment $a_{1}^{NLO}$ comes from contributions 
represented by the diagrams in Figs. \ref{figure2},\ref{figure3}.
We find
\begin{equation}
a_{1}^{NLO}=\frac{em_N^{2}}{6(4\pi f_{\pi })^2}
            \left[2h_{A}^{(2)}+g_{A}\left( h_{V}^{(0)}+\frac{4}{3}h_{V}^{(2)}
                  \right) \right]
            \ln \left( \frac{\mu ^{2}}{m_{\pi }^{2}}\right) 
            +\tilde{a}_{1}( \mu) ,
\end{equation}
where
\begin{equation}
\tilde{a}_{1}( \mu)=\tilde{a}_1%
+\frac{em^2_N}{6(4\pi f_\pi)^2}\left[ 2h^{(2)}_A%
 + g_A \left(h^{(0)}_V + \frac{4}{3}h^{(2)}_V \right)\right]
\left(\frac{1}{\varepsilon} +1 -\gamma -\frac{2}{3} +\ln 4\pi \right).
\end{equation}
Again, our result has the expected size and 
agrees
with Ref. \cite{mus+00}.
The term in $h_{A}^{(2)}$ agrees with Ref. \cite{Sav+93}.

\begin{figure}[tb]
\begin{center}
\epsfxsize=10cm
\centerline
{\epsffile{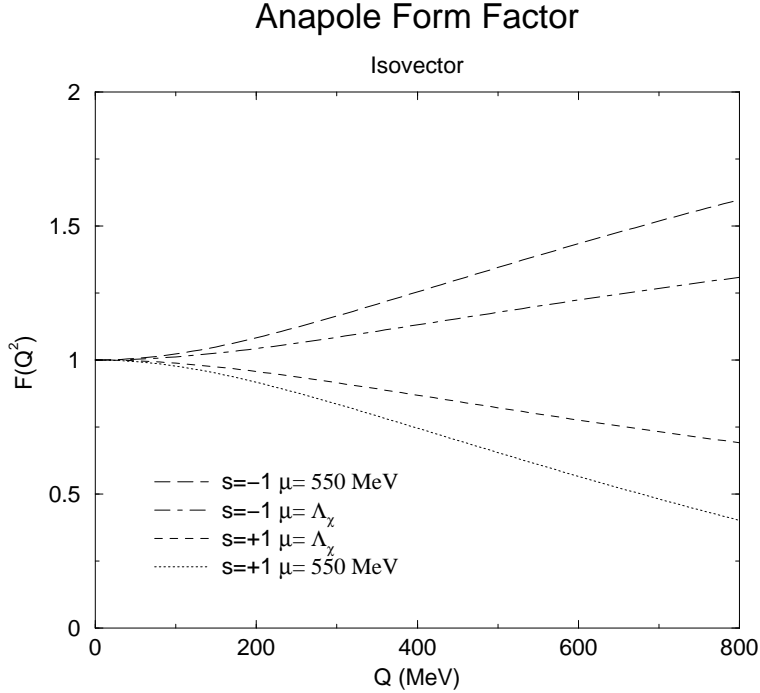}} 
\end{center}
\caption{The isovector anapole form factor $F^{NLO}_1$
as function of $Q$ in ChPT,
for a few reasonable values of parameters
expressed by the regularization scale 
$\mu$ that parametrizes the size of the
counterterm, and by  
$s$ that states the sign 
of the counterterm.}
\label{isovector}
\end{figure}

Contrary to the isoscalar part,
the isovector anapole form factor first appears in next-to-leading
order and reads
\begin{equation}
F_{1}^{NLO}( Q^{2}) =1
         -\frac{em_N^{2}}{9(4\pi f_{\pi })^{2}}\frac{1}{a_{1}^{NLO}} 
     \left[2h_{A}^{(2)}+g_{A}\left( h_{V}^{(0)}+\frac{4}{3}h_{V}^{(2)}
                  \right) \right]
           \left[ F^{NLO2}(Q^{2})-1\right].
\end{equation}
Again, for illustration we consider some representative values
of $\tilde{a}_{1}(\mu)$: 
$\tilde{a}_{1}( \Lambda _{\chi SB}) =0$,
$\tilde{a}_{1}( \Lambda _{\chi SB}) =
-2\alpha\ln \left( \frac{\Lambda _{\chi SB} ^{2}}{m_{\pi }^{2}}\right)$,
$\tilde{a}_{1}(550 {\rm MeV}) =0$,
$\tilde{a}_{1}(550 {\rm MeV}) =
-2\alpha\ln \left( \frac{(550 {\rm MeV}) ^{2}}{m_{\pi }^{2}}\right)$,
with $\alpha =\frac{em_N^{2}}{6(4\pi f_{\pi })^2}%
            \left[2h_{A}^{(2)}+g_{A}\left( h_{V}^{(0)}+\frac{4}{3}h_{V}^{(2)}%
                  \right) \right]$ and they all are summarized as
 \begin{equation}
F_{1}( Q^{2}) \simeq 1+s\frac{2}{3}\ln^{-1} \left( \frac{\mu ^{2}}{m_{\pi}^{2}}
        \right)\left[ F^{NLO2}( Q^{2})-1\right],\label{eq21}
\end{equation} 
where $s=- 1$ for $\tilde{a}_{1}( \mu) =0$ and $s= 1$ for 
$\tilde{a}_{1}(\mu) =-2\alpha\ln (\mu ^{2}/m_{\pi }^{2})$,
$\mu =0.55, 1.2$ GeV.
Fig. \ref{isovector} shows
$F_{1}^{NLO}( Q^{2}) $ for these four cases of $s$ and $\mu $.

The isovector mean square radius is 
\begin{equation}
\left\langle r_{1}^{2}\right\rangle ^{NLO}=
         \frac{1}{10m^{2}_\pi}\frac{em_N^{2}}
         {{a}_{1}^{NLO}(4\pi f_{\pi })^{2}}
         \left[2h_{A}^{(2)}+g_{A}\left( h_{V}^{(0)}+\frac{4}{3}h_{V}^{(2)}\right) 
         \right].
\end{equation}
Again, using the estimated form factor (\ref{eq21}) we have
\begin{equation}
\left\langle r_{1}^{2}\right\rangle ^{NLO} \simeq
          s\frac{6}{10m^{2}_\pi}
          \ln^{-1} \left( \frac{\mu}{m_\pi}\right)^2 .
\end{equation}

For $\mu =\Lambda_{\chi SB}$ one obtains
 $\left\langle r_{1}^{2}\right\rangle ^{NLO}=s(370 \ {\rm MeV})^{-2}$ and for  $\mu =550$ MeV, 
$\left\langle r_{1}^{2}\right\rangle ^{NLO}=s(298\ {\rm MeV})^{-2}$, where $s= \pm 1$.

We have thus for the first time calculated the momentum dependence
of the anapole form factor in next-to-leading order in ChPT.
Using dimensional analysis to estimate currently unknown
parameters, we see that the variation with momentum
is $\sim$ 20\% at $Q\sim 300$ MeV in both isoscalar and isovector channels.
The overall size of the anapole contribution to electron scattering
is thus likely not very different than that given by
the anapole moment itself. 
We can compare our result for
the isovector component to the forthcoming SAMPLE measurement.
The SAMPLE collaboration will extract an axial
contribution as seen by the electron, $G^e_A(0.1 {\rm MeV}^2)$
\cite{Bob}.
If this value is very different from the tree-level result,
it can only be assigned to the anapole form factor
if the parameters are substantially larger than
the naive dimensional expectation. 
Using our previous estimate,
\begin{equation}
2h^{(2)}_A +g_A \left( h^{(0)}_V + \frac{4}{3}h^{(2)}_V \right) = 
-\frac{6G_F (4\pi f_\pi)^2}{\eta F^{(1)}_A(Q^2)} 
\ln^{-1} \left( \frac{\Lambda_{\chi SB}}{m_\pi}\right)^2
\left[ G^e_A(Q^2) + G_A(Q^2) - G^s_A(Q^2)\right],
\end{equation}
where 
$\eta =8\sqrt{2}\pi \alpha/(1-4\sin^2\theta_W) = 3.45$,
$G_A(0)= 1.267$, $G^s_A(0)=-0.12$,
$G_A(Q^2)=G_A(0)/D(Q^2)$, 
$G^s_A(Q^2)= G^s_A(0)/D(Q^2)$,
$D(Q^2)=1+Q^2/M^2_A$,
and $M_A=1.061$ GeV.
For example, $G^e_A(0.1 {\rm MeV}^2)\sim 0.25$ requires
$2h^{(2)}_A +g_A ( h^{(0)}_V + \frac{4}{3}h^{(0)}_V) \sim
-10^{-5}$,
a hundred times larger in magnitude than dimensional analysis estimate.
This is very unlikely,
especially considering 
a recent estimate in the chiral quark model \cite{risk00}.

In any case, in the future, when parity-violating
pion-nucleon parameters are determined from other processes,
one can use the results reported here
to make firmer predictions for the anapole contribution
at various transferred momenta.

\vspace{1cm}
\noindent
{\large\bf Acknowledgements}

\noindent
We thank Bob McKeown and the group at the Kellogg Lab
for getting us interested in 
this problem, and Mike Musolf for discussions.  
CMM and JSV acknowledge fellowships from FAPESP 
(Brazil), grants 99/00080-5 and 99/05388-8. 
This research was supported in part by the US 
National Science Foundation. 


\vspace{1cm}
                  
\begin{thebibliography}{50}

\bibitem{Sample}  D.T. Spayde et al. (SAMPLE Collaboration),
{\it Phys. Rev. Lett.} {\bf 84} (2000) 1106;
B. Mueller et al. (SAMPLE Collaboration), 
{\it Phys. Rev. Lett.} {\bf 84} (1997) 3824. 

\bibitem{SampleII} E. Beise and  M. Pitt (co-spokesperson), MIT-Bates experiment 94-11 .

\bibitem{Zei58}  Ya.B. Zel'dovich, {\it Sov. Phys. JETP} {\bf 6} (1958) 1184; 
{\it Sov. Phys. JETP} {\bf 12} (1961) 777.

\bibitem{Wood97}  C.S. Wood et al., {\it Science} {\bf 275} (1997) 1759.

\bibitem{Mus+91}  M.J. Musolf and B.R. Holstein, 
{\it Phys. Rev. } {\bf D43} (1991) 2953; 
W.C. Haxton, E.M. Henley, and M.J. Musolf, 
{\it Phys. Rev. Lett.} {\bf 63} (1989) 949.

\bibitem{Sav+93}  D.B. Kaplan and M.J. Savage, {\it Nucl. Phys. }{\bf A556}
(1993) 653.

\bibitem{Bob}  R.D. McKeown, in {\it Parity Violation in Atoms and
Polarized Electron Scattering}, ed. by B. Frois and M.A. Bouchiat, 
World Scientific (1999), p. 423.

\bibitem{Happex}  K.A. Aniol et al. (Happex Collaboration), 
{\it Phys. Rev. Lett.} {\bf 82} (1999) 1096.

\bibitem{G0}  G0 Collaboration, www.npl.uiuc.edu/exp/G0/G0Main.html.

\bibitem{tito}  T.M. Ito (Spokesperson), MIT-Bates experiment 00-004.

\bibitem{mei} V. Bernard, N. Kaiser, and U.-G. Mei{\ss}ner, 
{\it Int. J. Mod. Phys.} {\bf E4} (1995) 193.

\bibitem{bira}  U. van Kolck, {\it Prog. Part. Nucl. Phys. }{\bf 43}
(1999) 337.

\bibitem{mus+00} S.L. Zhu, S.J. Puglia, B.R. Holstein, M.J.R. Musolf, 
{\tt hep-ph/0002252}.

\bibitem{Sav+98} M.J. Savage and R.P. Springer, {\tt nucl-th/9907069}. 

\bibitem{bira1}  C.M. Maekawa and U. van Kolck, {\it Phys. Lett.} {\bf B478}
(2000) 73.

\bibitem{risk00} D.O. Riska, {\tt hep-ph/0003132}.


\end{thebibliography}
\end{document}